\documentclass[%
preprint, 
superscriptaddress,
 amsmath,amssymb,
 aps,
prfluids,
 onecolumn,
]{revtex4-1}

\usepackage{graphicx}
\usepackage{dcolumn}
\usepackage{bm}

\usepackage{xcolor}
\usepackage{color}

\newcommand{\dd}{\ensuremath{\mathrm{d}}}

\newcommand{\f}[2]{\ensuremath{\frac{#1}{#2}}} 

 \mathchardef\mhyphen="2D

\newcommand\Pran{\mbox{\textit{Pr}\hspace{0.02cm}}} 
\newcommand\Sca{\ensuremath{{\gamma_{eff}}}} 

\definecolor{myred2}{RGB}{179, 13, 13}
\definecolor{myred}{RGB}{255, 0, 0}
\definecolor{myblue2}{RGB}{0, 0, 230}
\definecolor{myblue}{RGB}{0, 0, 255}
\definecolor{mygreen}{RGB}{0, 102, 0}
\definecolor{myorag}{RGB}{246, 150, 38}
\definecolor{mypurp}{RGB}{73,0,146}
\definecolor{myyel}{RGB}{235,235,10}
\definecolor{LGrey}{rgb}{.5,.5,.5}
\definecolor{change}{rgb}{0,0,1}

\makeatletter
\newif\ifpreprintoption
\@ifclasswith{revtex4-1}{preprint}{\preprintoptiontrue}{\preprintoptionfalse}
\makeatother

\begin{document}


\title{Heat transfer in rough-wall turbulent \textcolor{black}{thermal convection} in the ultimate regime}

\author{Michael MacDonald}
	\email{michael.macdonald@unimelb.edu.au}
	\altaffiliation[now at ]{Jet Propulsion Laboratory, California Institute of Technology, Pasadena, CA 91109, USA}
	\affiliation{Department of Mechanical Engineering, University of Melbourne, Victoria 3010, Australia}
\author{Nicholas Hutchins}
	\affiliation{Department of Mechanical Engineering, University of Melbourne, Victoria 3010, Australia}
\author{Detlef Lohse}
	\affiliation{ Physics of Fluids Group, MESA+ Institute, J. M. Burgers Center for Fluid Dynamics and Max Planck Center Twente, \\ University of Twente, P.O. Box 217, 7500AE Enschede, The Netherlands}
	\affiliation{ Max Planck Institute for Dynamics and Self-Organization, 37077 G{\"o}ttingen, Germany}
\author{Daniel Chung}
	\affiliation{Department of Mechanical Engineering, University of Melbourne, Victoria 3010, Australia}

\date{\today}

\begin{abstract}

Heat and momentum transfer in wall-bounded turbulent flow, 
coupled with the effects of wall-roughness, is one of the outstanding
questions in turbulence research.
In the standard Rayleigh--B\'enard problem for natural thermal convection,
it is notoriously difficult to  reach the so-called ultimate regime 
 in which the near-wall boundary layers are turbulent. 
Following the analyses proposed by 
Kraichnan [Phys.~Fluids {\bf 5}, 1374--1389 (1962)]
and
 Grossmann \& Lohse [Phys.~Fluids {\bf 23}, 045108 (2011)],
we instead utilize recent direct numerical simulations of
  {\it forced} convection over a rough wall in a minimal channel
[MacDonald, Hutchins \& Chung, J.~Fluid Mech.~{\bf 861}, 138--162 (2019)]
to directly study these turbulent boundary layers.
We focus on the heat transport
(in dimensionless form, the Nusselt number $Nu$) or equivalently the 
heat transfer coefficient (the Stanton number $C_h$).
Extending the analyses of Kraichnan and Grossmann \& Lohse,
we assume logarithmic temperature profiles with a roughness-induced shift
to predict an effective scaling of $Nu \sim Ra^{0.42}$,
where $Ra$ is the dimensionless temperature difference,
corresponding to $C_h \sim Re^{-0.16}$, where $Re$ is the centerline Reynolds number.
This is pronouncedly  different from the skin-friction coefficient $C_f$,
which in the fully rough turbulent regime is independent of $Re$, due to the dominant pressure drag. 
In rough-wall turbulence
 the absence of the analog to pressure drag
in the temperature advection equation is the origin for the very different scaling properties
of the 
heat transfer as compared to the momentum transfer. 
This analysis suggests that, unlike momentum transfer, the asymptotic ultimate regime, where $Nu\sim Ra^{1/2}$,
will never be reached for heat transfer at finite $Ra$.

\end{abstract}

\maketitle


\section{Introduction}

Heat transfer in wall-bounded turbulent flow  is one of the outstanding problems in turbulence, both from a fundamental and an applied point of view. 
The canonical system to study it is Rayleigh--B\'enard (RB) convection \cite{ahl09,loh10,chi12}, i.e., the flow 
in a container heated from below and cooled from above. Here the key question is: how does the 
heat transfer (in dimensionless form, the Nusselt number $Nu$) scale with the temperature difference
between the top and bottom walls   (in dimensionless form, the Rayleigh number $Ra$)? And how does this 
scaling change with wall-roughness? 
For smooth walls in the so-called classical regime, in which the boundary layers (BLs)
are of Prandtl--Blasius (i.e. basically laminar) type,  the dependencies  are
reasonably understood along the unifying theory
of thermal convection \cite{gro00,gro01,ahl09,ste13}.
However, the situation is much 
less clear in the so-called ultimate regime,
in which the BLs become turbulent \cite{kra62,spi71,gro11,gro12} and the heat transfer is thus enhanced. 
In this regime 
Kraichnan \cite{kra62} predicted that $Nu\sim Ra^{1/2}[\log(Ra)]^{-3/2}$. 
Later, Grossmann \& Lohse \cite{gro11,gro12} used logarithmic velocity and temperature profiles to quantify the logarithmic correction term.
Beyond the transition, which for gases was predicted \cite{gro01}
 to occur around $Ra \sim 10^{14}$, 
both predictions imply an effective scaling $Nu \sim Ra^\Sca$ with 
$\Sca\approx0.38$.
As $Ra\rightarrow\infty$, the logarithmic correction terms become negligible and the flow approaches the so-called asymptotic ultimate regime where $Nu\sim Ra^{1/2}$. This asymptotic ultimate 
regime implies that viscosity and thermal diffusivity effects have a negligible impact on the flow. In contrast to  \cite{kra62,spi71,gro11}, 
Owen and Thompson \cite{Owen63} proposed that the asymptotic ultimate (and upper bound 
\cite{how63,bus69,doe96})
scaling exponent 1/2 is never
achieved.

Whether and when  the transition to the ultimate regime indeed occurs, and to what turbulent state, 
has been hotly debated in the community.
While Ahlers, Bodenschatz, and He 
 experimentally found such a transition around $Ra\sim10^{14}$
\cite{he12,he12a,ahl14}, Chavanne, Roche {\it et al.} \cite{cha97,cha01,roc10} observed it
 at lower $Ra\sim10^{11} - 10^{12}$ and  others still do not find such a transition at all \cite{nie00,nie10,urb14}. 
To clarify this question, major numerical efforts are undertaken to solve the underlying 
Boussinesq  equations in this large $Ra$  regime. While in 3D the required computational power is presently
prohibitive, in 2D the onset of such a transition has recently been observed around $Ra\sim10^{13}$ 
\cite{zhu18b,kru18}, namely in the effective scaling of $Nu(Ra)$ and in the structure of the BLs, changing towards a 
logarithmic profile in the ultimate regime. 

To trigger the onset of the ultimate regime, i.e.~the transition from a laminar-type BL to a turbulent one,
various wall-roughness elements have been employed, both in experiments  
\cite{she96,cil99,du00,roc01,qiu05,tis11,sal14,wei14,xie17,Tummers19} and in numerical 
simulations \cite{str06,shi11,wag15,zhu17b}.  In general, these efforts have led to an enhanced $Nu$ versus $Ra$ scaling
in some intermediate $Ra$ regime, and in limited $Ra$ regimes even an effective $Nu$ versus $Ra$ scaling exponent of 1/2 can
be achieved. 
For large $Ra$ (but still below the onset of the ultimate regime) the effective scaling exponent settles back to a value
close to the one in the classical  regime \cite{zhu17b},
 as then the  thermal sublayer is thinner than the roughness elements, and starts to conform to the roughness topography.
Only for even larger $Ra$ -- hitherto not yet achieved in rough-wall RB 
flow -- is the transition towards a turbulent BL throughout, and enhanced $Nu$ versus $Ra$,  expected. 

To address the question of heat transport in smooth and rough-wall RB convection in the ultimate regime,
 in this paper we will 
assume the hypothesis proposed by Kraichnan \cite{kra62} and Grossmann \& Lohse \cite{gro11,gro12}
that the boundary layers are turbulent with logarithmic profiles. 
This allows us to employ our understanding
of heat transfer in smooth and in particular
  rough-wall fully 
  turbulent forced convection channel flow.  
   The advantage of such flow is that the driving is mechanically supplied (namely by shear),
  which is much more efficient than the thermal driving in RB flow. Therefore, in numerical simulations
 the transition to turbulence in the boundary layers -- manifesting itself in a logarithmic velocity profile 
 -- can easily be achieved \cite{smi11,smi13}.
 Such turbulent boundary layers with a logarithmic velocity profile also exist in the shear-driven Taylor--Couette (TC)
 flow \cite{gro16}, which is viewed as the ``twin'' of RB flow \cite{bus12}. For that flow, indeed the 
 ultimate regime with the corresponding Nusselt number $Nu_\omega$ (the dimensionless angular velocity
 transport \cite{eck07b}) scaling $Nu_{\omega}\propto Ta^{0.38}$ (where the Taylor number $Ta$ is the dimensionless 
 mechanical driving strength) can be achieved both in experiments  and in numerical simulations, see the 
 review article \cite{gro16}. In TC flow with a rough wall, even the asymptotic ultimate regime $Nu_\omega\propto Ta^{1/2}$ can be achieved, both experimentally \cite{cad97,ber03,zhu18} and numerically \cite{zhu18}.
 This regime corresponds to fully rough pipe or channel flow in which the  friction 
 factor becomes Reynolds number independent
 \cite{Nikuradse33,Hama54,Jimenez04,zhu18}.
 The reason is that in this regime the drag is determined by the pressure drag, and shear (viscous) drag hardly plays a role. 
However, this dominant pressure drag also implies that the analogy between heat transfer in RB and angular velocity transport  in TC breaks down
 for roughness, due to the lack of a pressure-like component in the temperature advection equations  \cite{zhu17b}.
 The quantitative effect of roughness on the heat transfer in RB flows,  despite this qualitative understanding,
 is therefore not  well understood, and we will
 address it here using forced convection channel flow. 
Note that, like RB and TC flows, numerical simulations of closed channel (and pipe) flows employing periodic boundary conditions in the flow direction also enjoy exact energy balances \cite{eck07a}.
 
In the present work, we will use the recent rough-wall turbulent forced convection  results from \cite{mac19}
as a model for the near-wall shear-dominated turbulent boundary layers observed in high $Ra$ natural convection flows, as envisioned by Kraichnan \cite{kra62}.
We will therefore seek to quantify and explain the effect of roughness on the scaling exponent of the Nusselt number
in the ultimate regime. 
This involves extending the analysis of Ref.~\cite{gro11} for smooth-wall ultimate RB flow with logarithmic velocity and temperature profiles, by
quantifying the shift in the profiles  induced by the roughness.

Briefly we summarize the forced convection direct numerical simulations (DNSs) of Ref.~\cite{mac19},
in which buoyancy forces were neglected so that temperature was a passive scalar.
Periodic boundary conditions were employed in the streamwise ($x$) and spanwise ($y$) directions and no-slip, impermeability and isothermal ($\theta_w=0$) conditions were applied to the top and bottom walls, with $z$ denoting the wall-normal (vertical) direction. 
A body forcing to the momentum equation was used to drive the flow at constant bulk velocity through the channel. 
An internal heating body force to the energy equation was used for temperature, representing a hot fluid being cooled by the walls. 
The Prandtl number was set to that of air at room temperature, $Pr\equiv\nu/\kappa=0.7$,
where $\nu$ is the kinematic viscosity and $\kappa$ is the thermal diffusivity.
 Different friction Reynolds numbers, $Re_\tau=U_\tau h/\nu$ were simulated with $395\lesssim Re_\tau\lesssim1680$, where $U_\tau$ is the friction velocity and $h$ is the channel half height, defined for the rough-wall flow to be distance between the channel center and the roughness mean height, corresponding to the hydraulic half height \citep{Chan15}. 
 Three-dimensional sinusoidal roughness with semi-amplitude of either $k=h/18$ or $k=h/36$ and wavelength $\lambda_x=\lambda_y=\lambda\approx7.07k$ was applied to both the bottom and top walls. As the flow speed and friction Reynolds number increases, the roughness Reynolds number $k^+=kU_\tau/\nu$  increases towards the fully rough regime.
Superscript $+$ indicates non-dimensionalization on $\nu$, $U_\tau\equiv\sqrt{\tau_w/\rho}$ and the friction temperature $\Theta_\tau\equiv [q_w/(\rho c_p)]/ U_\tau$,  
 $\tau_w$ and $q_w$ being the temporally and spatially averaged momentum and heat fluxes at the wall, $\rho$ the fluid density and  $c_p$ the specific heat at constant pressure.
The minimal-span channel for rough wall flows was used \citep{Chung15,MacDonald17}, in which the spanwise domain width is purposely very narrow and only the near-wall turbulent flow is captured up to a critical height $z_c\approx0.4L_y$, where $L_y$ is the channel span.  
 Smooth-wall channel simulations with matched channel domain sizes were also  conducted, to ensure that the differences between the smooth- and rough-wall flows were due to the roughness alone and not the channel span. 
\setlength{\unitlength}{1cm}
\begin{figure*}
\centering
\includegraphics[width=\textwidth]{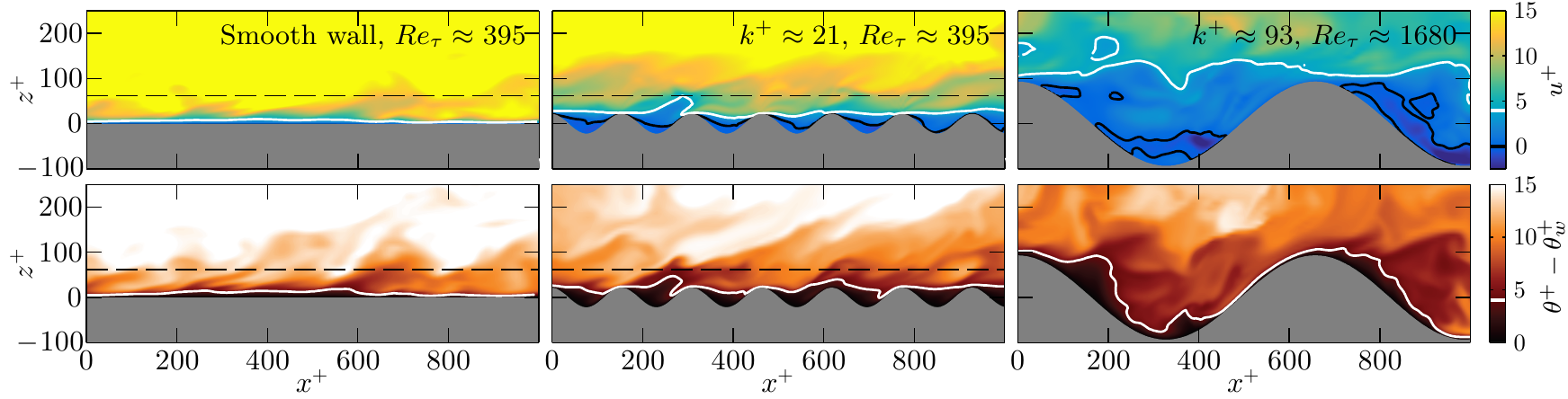}
\vspace{-2.0\baselineskip}
\caption{
Instantaneous steamwise velocity (top) and temperature (bottom)
for smooth-wall (left), transitionally rough  (center) and fully rough (right) flows, from forced convection DNS \cite{mac19}.
The horizontal dashed line shows the minimal channel critical height, $z_c=0.4L_y$. The black contour line shows zero velocity (recirculation), while the white contour line shows the value $u^+=4$ and $\theta^+-\theta_w^+=4$, to highlight the viscous and thermal diffusive sublayers. Flow is from left to right.
 }
	\label{fig:instTemp}
\end{figure*}

Fig.~\ref{fig:instTemp} shows instantaneous snapshots of the streamwise velocity and fluid temperature from the simulations of \cite{mac19}. A white contour line for $u^+=4$ and $\theta^+-\theta_w^+=4$ has been selected to provide an indication of the viscous and thermal diffusive sublayers.
The smooth-wall flow 
 produces thin viscous and thermal sublayers close to the wall that appear similar, although the thermal sublayer is slightly larger due to the Prandtl number being less than unity.
The roughness in the transitionally rough regime ($k^+\approx 21$) 
produces much thicker sublayers and, like the smooth wall, appear similar between  the velocity and temperature fields.
The final (right most) panel shows much larger roughness that is nominally fully rough ($k^+\approx93$).
The selected contour line of $u^+=4$ resides mostly above the roughness crests, indicating the fluid below the level of the roughness crests is nearly stationary due to the increased dominance of pressure drag. 
Conversely, the thermal diffusive sublayer is thin and closely follows the roughness geometry; it appears more like that of the smooth wall if the wall were contorted to match the roughness geometry.

\section{Turbulent Boundary Layers in the Rough-Wall Ultimate Regime}

The turbulent boundary layers observed in high $Ra$ (ultimate regime) natural convection flows are characterized by
local buoyancy forces that are much smaller than the shear forces. This leads to
mean velocity and temperature profiles that are logarithmic functions of distance from the wall 
 \cite{gro11}, the same as in forced convection flows \cite{Kader81,Kawamura99,Pirozzoli16}, given as 
\begin{eqnarray}
\label{eqn:logU} 
\overline{u}^+		&\equiv& \f{\overline{u}}{U_\tau} = \frac{1}{\kappa_m}\log\left(z^+\right)+A_m-\Delta U^+(k^+), \\
\label{eqn:logT}
\overline{\theta}^+-\overline{\theta}_w^+&\equiv& \f{\overline{\theta}-\overline{\theta}_w}{\Theta_\tau} = \frac{1}{\kappa_h}\log(z^+)+A_h(Pr)-\Delta \Theta^+(k^+)
\end{eqnarray}
where $\kappa_m\approx0.4$ is the von K{\'a}rm{\'a}n constant, which is slightly larger for heat transfer with $\kappa_h\approx0.46$ due to the turbulent Prandtl number (the ratio of momentum and heat transfer eddy diffusivities) being below unity \cite{Yaglom79,Pirozzoli16}.
As in \cite{mac19} we take the smooth-wall offsets  to be $A_m\approx5.0$ and $A_h(Pr=0.7)\approx3.2$. 
The enhanced skin friction and heat transfer of roughness  lead to a downwards shift of these logarithmic profiles, represented by the roughness function, $\Delta U^+$ \cite{Hama54}, and temperature difference, $\Delta \Theta^+$ \cite[][]{Yaglom79,Miyake01,Leonardi07tsfp,mac19}.
These two quantities are a flow property of a given rough surface and vary with the roughness Reynolds number.
Evaluating Eqs.~(\ref{eqn:logU}--\ref{eqn:logT}) at the middle of an RB cell, $z=L/2$, yields
\begin{eqnarray}
\label{eqn:U_imp}
U^+	\equiv \overline{u}^+\left(z=\f{L}{2}\right)		&=& \frac{1}{\kappa_m}\log\left(\f{\f{1}{2}Re}{U^+}\right)+A_m-\Delta U^+\left(\f{k}{L} \f{Re}{U^+}\right), \\
\label{eqn:T_imp}
\Theta^+\equiv \overline{\theta}^+\left(z=\f{L}{2}\right) -\overline{\theta}_w^+	&=& \frac{1}{\kappa_h}\log\left(\f{\f{1}{2}Re}{U^+}\right)+A_h(Pr)-\Delta \Theta^+\left(\f{k}{L} \f{Re}{U^+}\right),
\end{eqnarray}
where the Reynolds number $Re=UL/\nu$.
Eqs.~(\ref{eqn:U_imp}) and (\ref{eqn:T_imp}) thus describe $U^+$ and $\Theta^+$ for a given flow state governed by $Re$ and $Pr$ and by the relative roughness $k/L$,
provided $\Delta U^+$ and $\Delta \Theta^+$ are known.

We define the skin-friction coefficient as $C_f\equiv \tau_w/(\tfrac{1}{2}\rho U^2)=2/U^{+2}$
and
heat-transfer coefficient (Stanton number) as $C_h \equiv q_w/(\rho c_p U) = 1/(U^+ \Theta^+)$.
The temperature profiles from the top and bottom walls must match at the centerline so that $2\Theta=\Delta$,
where $\Delta$ is the driving temperature difference in RB domains and
thus we define the Nusselt number as $Nu = \Pran Re/(U^+\Delta^+)=\Pran Re C_h/2$.
In order to describe the rough-wall $C_f$, $C_h$ and $Nu$ as a function of Reynolds number,
we therefore require knowledge of $\Delta U^+$ and $\Delta \Theta^+$.
Note that for the smooth wall case ($\Delta U^+=\Delta \Theta^+=k/L=0$), Eq.~(\ref{eqn:U_imp}) yields the implicit Prandtl--von K{\'a}rm{\'a}n logarithmic skin-friction law, which can be solved using Lambert's $\mathcal{W}$-function with $U^+=(1/\kappa_m)\mathcal{W}\left(Re\hspace{0.04cm}\kappa_m\exp\left(A_m\kappa_m\right)/2\right)$. Grossmann \& Lohse \cite{gro11} obtained the smooth-wall ultimate-regime Nusselt number scaling exponent of $\Sca\approx0.38$ using this result.

\setlength{\unitlength}{1cm}
\begin{figure}
\centering
	\includegraphics{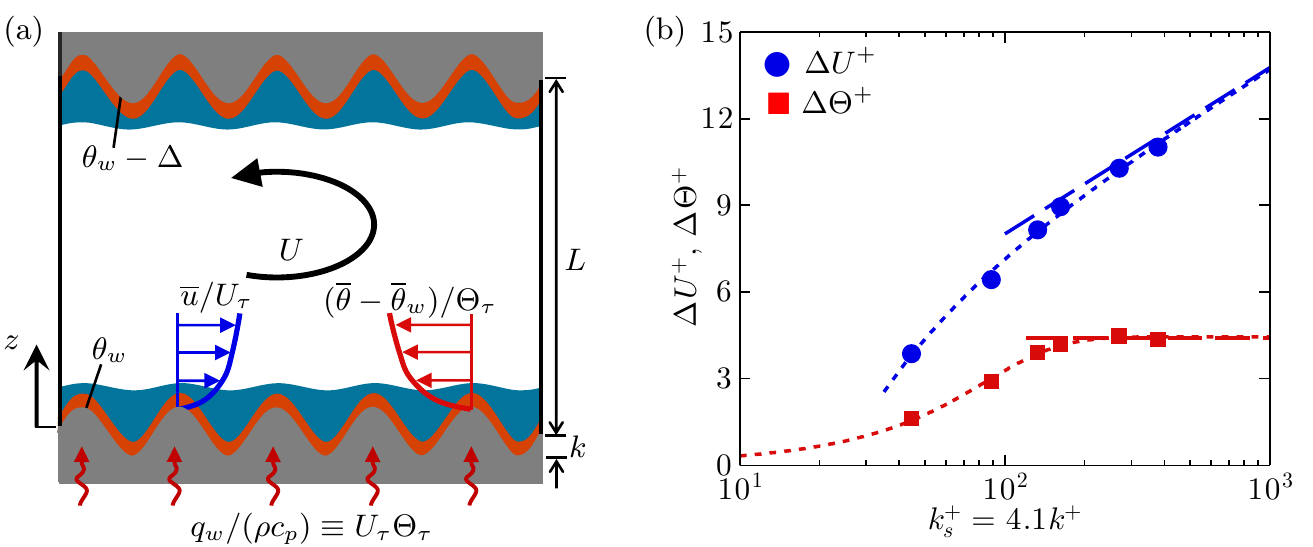}
	\vspace{-0.7\baselineskip}
\caption{
(a) Sketch of the Rayleigh--B\'enard system with roughness.
Under fully rough conditions, the viscous sublayer (blue) is larger than
the thin, roughness-conforming thermal sublayer (orange) 
(Fig.~\ref{fig:instTemp}).
The near-wall velocity and temperature profiles
are logarithmic in the ultimate regime \cite{kra62,gro11} but, relative to a smooth wall, are shifted by $\Delta U^+$ and $\Delta \Theta^+$, respectively, due to the roughness (Eqs.~\ref{eqn:logU}--\ref{eqn:logT}).
(b)
Roughness function, $\Delta U^+$ (blue),
and
temperature difference, $\Delta \Theta^+$ (red),
 as a function of the equivalent sand-grain roughness Reynolds number, $k_s^+\approx 4.1 k^+$. 
Symbols are the forced convection DNS data \cite{mac19};
dotted lines are curve fits to the DNS data (see text);
dashed lines are the fully rough asymptotes.
}
	\label{fig:DUDT}
\end{figure}

Fig.~\ref{fig:DUDT}(a) shows a sketch of a sinusoidal rough-wall RB domain.
We can obtain the roughness function $\Delta U^+$ and temperature difference $\Delta \Theta^+$ from the turbulent forced convection system of \cite{mac19}.
These two quantities are shown in Fig.~\ref{fig:DUDT}(b), as a function of the equivalent sand-grain roughness Reynolds number, $k_s^+\approx4.1k^+$.
This scaling guarantees a collapse of the roughness function with that of Nikuradse's sand-grain data in the fully rough regime (here $k_s^+\gtrsim150$),
where $\Delta U_{FR}^+=(1/\kappa_m)\log(k_s^+)+A_m-C_n$ (blue dashed line), with $C_n\approx8.5$ being Nikuradse's constant \cite{Nikuradse33,Schlichting36}.
Within the asymptotic fully rough regime, viscous effects are negligible and the pressure (or form) drag is dominant \citep{Schultz09,Busse17,zhu18}.
Note that $k_s$  must be determined dynamically for a given rough surface and is not a simple geometric length scale of the roughness.
The temperature difference, meanwhile, is tending towards a constant value of $\Delta \Theta_{FR}^+\approx 4.4$ in the fully rough regime (red dashed line). 
Like $k_s$, the exact value of $\Delta \Theta_{FR}^+$ is a dynamic parameter that is likely to be roughness dependent. Ultimately however, with this information, in the fully rough regime Eqs.~(\ref{eqn:U_imp}) and (\ref{eqn:T_imp}) become
\begin{eqnarray}
\label{eqn:UbFR_C}
 U_{{FR}}^+ &=& C_n-\frac{1}{\kappa_m}\log\left(\frac{k_s}{L/2}\right), \\
\label{eqn:TbFR}
\Theta_{{FR}}^+	&=& \frac{1}{\kappa_h}\log\left(\frac{\frac{1}{2}Re}{U_{{FR}}^+}\right)+A_h-\Delta \Theta_{FR}^+.
\end{eqnarray}
That is, the friction-normalized centerline velocity is constant and only depends on the relative roughness $k_s/L$, while the centerline friction-normalized temperature remains dependent on the Reynolds number.

The dotted lines in Fig.~\ref{fig:DUDT}(b) are curve fits to the DNS data. 
For the roughness function, 
we use the fit from Ref.~\cite{Monty16} with $\Delta U_{fit}^+\approx (1/\kappa_m)\log(k_s^+)+A_m-C_n-(a/k_s^+)^b$, where $a\approx89.3$ and $b\approx1.12$. While the fitting constants are different to \cite{Monty16} due to the different roughness geometries, this function correctly tends towards the fully rough asymptote for large $k_s^+$. However, the function quickly reaches zero at $k_s^+\approx 25$, much more rapidly than for sinusoidal roughness or  sand-grain roughness (see figure 5 of \cite{mac19}). We therefore only use $\Delta U_{fit}^+$ for $k_s^+\gtrsim35$.
 For the temperature difference, a sigmoid function across the entire range of $k_s^+$ is used, with $\Delta \Theta_{fit}^+\approx B+K/[1+\exp(c+d k_s^+)]$ where $B\approx-1.66$, $K\approx6.11$, $c\approx0.97$ and $d\approx-0.0239$. This function correctly goes to zero for small $k_s^+$ (i.e. a smooth wall) and tends towards $\Delta \Theta_{FR}^+\approx4.4$ in the limit of $k_s^+\rightarrow\infty$ (i.e. the fully rough regime). 
 We can therefore obtain $C_f$, $C_h$ and $Nu$ numerically using $\Delta U_{fit}^+$ and $\Delta \Theta_{fit}^+$ with Eqs.~(\ref{eqn:U_imp}--\ref{eqn:T_imp}), although the functional form of $\Delta U_{fit}^+$ precludes an analytical solution for $U^+$.

\section{Effective heat-transfer scaling in the ultimate regime}
\label{sect:scaling}

\setlength{\unitlength}{1cm}
\begin{figure*}
\centering
\includegraphics[width=\textwidth]{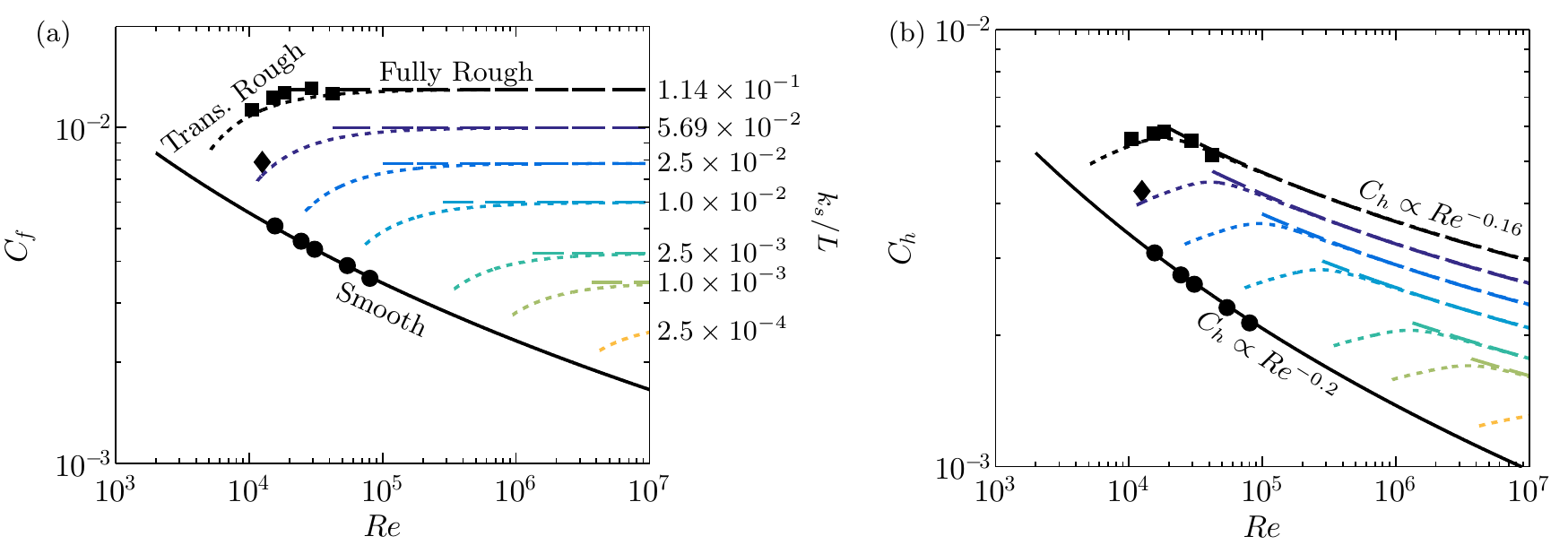}
\vspace{-0.7\baselineskip}
\caption{
(a)
Skin-friction coefficient and
(b)
 heat-transfer coefficient (Stanton number) against  Reynolds number. 
DNS data \cite{mac19} shown by symbols:
circles, smooth wall;
squares, rough wall $k_s/L=1.14\times10^{-1}$;
diamond, $k_s/L=5.69\times10^{-2}$.
Line styles: 
solid, smooth wall  (Eqs.~(\ref{eqn:U_imp}--\ref{eqn:T_imp}) with $\Delta U^+=\Delta \Theta^+=0$);
dotted, transitionally rough regime  (Eqs.~(\ref{eqn:U_imp}--\ref{eqn:T_imp}) with $\Delta U_{fit}^+$ and $\Delta \Theta_{fit}^+$ curve  fits from Fig.~\ref{fig:DUDT}b);
dashed, fully rough regime (Eqs.~\ref{eqn:UbFR_C}--\ref{eqn:TbFR}).
Different colors (in both figures) correspond to different relative roughnesses $k_s/L$, with legend in (a).
}
	\label{fig:CfCh}
\end{figure*}

Fig.~\ref{fig:CfCh}(a) and (b) shows the skin-friction and heat-transfer coefficients as a function of Reynolds number.
Here, the solid lines show the smooth-wall $C_f$ and $C_h$  calculated using the logarithmic velocity and temperature profiles (Eqs.~(\ref{eqn:U_imp}--\ref{eqn:T_imp}) with $\Delta U^+=\Delta \Theta^+=0$).
The dotted lines correspond to the rough-wall $C_f$ and $C_h$ calculated using Eqs.~(\ref{eqn:U_imp}--\ref{eqn:T_imp}) with the curve fits  $\Delta U_{fit}^+$ and $\Delta \Theta_{fit}^+$  from Fig.~\ref{fig:DUDT}(b),
while
the dashed lines show the asymptotic fully rough regime (Eqs.~\ref{eqn:UbFR_C}--\ref{eqn:TbFR}). The relative roughness ratio $k_s/L$ is varied, given by the different colors, where it is assumed $\Delta U^+$ and $\Delta \Theta^+$ are independent of $k_s/L$.
The symbols are the DNS data \cite{mac19}, assuming that the channel centerline $h$ is equal to the middle of the RB cell, $L/2$.
For the smooth-wall flow, both of these coefficients monotonically reduce with Reynolds number at the same rate.
Roughness enhances momentum and thermal transfer, leading to an increase in these coefficients relative to the smooth wall.
In the fully rough regime ($Re\gtrsim 2\times 10^4$ for $k_s/L=1.14\times10^{-1}$, black dashed  line) the skin-friction coefficient becomes constant with Reynolds number, with $C_f\approx0.013$.
This indicates that the viscous effects are negligible and the momentum transfer has attained an asymptotic state \cite{zhu17b}.
Conversely, the heat-transfer coefficient reduces with Reynolds number in the fully rough regime, similar to the smooth wall, and does not appear to reach any asymptotic state.
At the present Reynolds numbers, Eqs.~(\ref{eqn:U_imp}--\ref{eqn:TbFR}) predict $C_h\propto Re^{-0.2}$ for the smooth wall \cite[e.g.][]{Kays05}, while for the rough wall with $k_s/L=1.14\times10^{-1}$ it scales as $C_h\propto Re^{-0.16}$.

\setlength{\unitlength}{1cm}
\begin{figure*}
\centering
	\includegraphics[width=\textwidth]{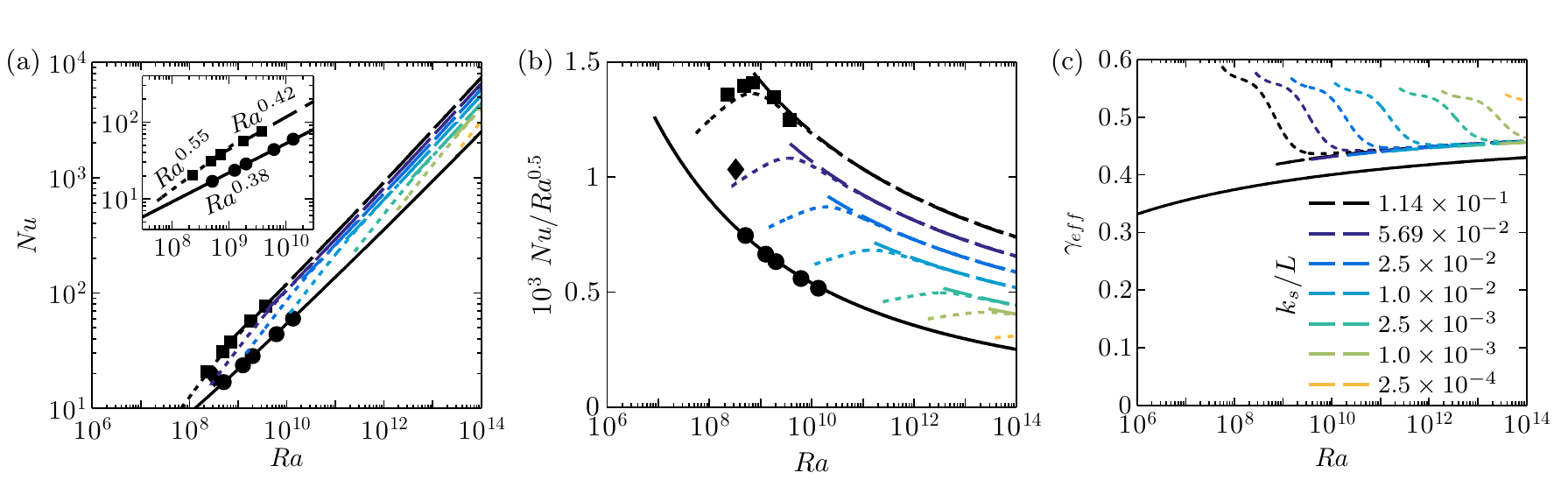}
	\vspace{-1.5\baselineskip}
\caption{
(a)
Nusselt number, $Nu$,
(b)
compensated $Nu/Ra^{0.5}$ and
(c) 
predicted (from log equations) effective scaling exponent, $\Sca$, in $Nu\propto Ra^{\Sca}$,
as as function of the Rayleigh number.
The inset in (a) highlights just the forced convection DNS data \cite{mac19}.
Symbols and line styles are the same as Fig.~\ref{fig:CfCh}(a).
}
	\label{fig:NuRa}
\end{figure*}

Ultimately we would like to know the dependency of the Nusselt number on the Rayleigh number, $Nu\sim Ra^{\Sca}$.
However, the log equations (Eqs.~\ref{eqn:U_imp}--\ref{eqn:TbFR}) as well as the forced convection DNS data are functions of the Reynolds number,
requiring an assumption of the form  $Re= A\cdot Ra^\beta$ in order to determine the Rayleigh number dependency.
Grossmann \& Lohse \cite{gro11} made the same assumption in their work,
where they used $Re = 0.346\cdot  Ra^{0.443}$, corresponding to boundary layers that were not yet turbulent.
As we are explicitly dealing with turbulent (logarithmic) boundary layers, we use 
the ultimate-regime scaling of $\beta=1/2$ \cite{gro11}.
More recently, \citeauthor{Shishkina17} \cite{Shishkina17} derived this relation with $\beta=1/2$ from the Prandtl BL equations with very weak assumptions.
We take the coefficient $A=0.7$, which results in $Nu(Ra=10^{14})\approx2500$ for the smooth wall, in agreement the laboratory experiments of Ref.~\cite{he12a}.
The exact choice of $A$ and $\beta$ will alter the absolute values of $Nu$ and $\Sca$ for a given $Ra$, however we emphasize that the assumption made here does not alter the main conclusions of this paper.

Figure \ref{fig:NuRa}(a) shows the Nusselt number as a function of the Rayleigh number.
The inset highlights just the DNS data,
where the smooth-wall forced convection data (circles) have an effective scaling exponent of $\Sca\approx0.38$ (solid black line),
matching that observed in ultimate RB convection \cite{gro11}, as expected.
Note that the mechanically supplied shear of forced convection corresponds to a very strong wind in RB convection, explaining the relatively
low $Ra$ values in the inset of Fig.~\ref{fig:NuRa}(a).
A realistic RB flow in the ultimate regime typically has higher $Ra$ values due to a weaker wind (i.e., a lower Reynolds number, or coefficient $A$ in the assumed $Re-Ra$ relationship above).
The transitionally rough regime, meanwhile, has an enhanced exponent of  $\Sca\approx0.55$ shown by the dotted lines (similar to the rough-wall  RB experiments by \cite{xie17}),
however in the asymptotic fully rough regime the scaling reduces towards $\Sca\approx0.42$.
Fig.~\ref{fig:NuRa}(b) shows the compensated form, $Nu/Ra^{1/2}$, demonstrating that while the transitionally rough regime has a scaling near $\Sca\approx0.55$, the fully rough regime clearly has a reduced scaling exponent.

 Figure~\ref{fig:NuRa}(c) shows the effective scaling exponent, $\dd \log_{10}(Nu)/\dd \log_{10}(Ra)$,  computed using the log-law formulas for smooth-wall and fully rough flows (Eqs.~\ref{eqn:U_imp}--\ref{eqn:TbFR}). This is also done using the curve fits of $\Delta U_{fit}^+$ and $\Delta \Theta_{fit}^+$ (Fig.~\ref{fig:DUDT}b) in Eqs.~(\ref{eqn:U_imp}--\ref{eqn:T_imp}), shown by the dotted lines.
 This figure is similar to figure 2a of Ref.~\cite{gro11}, where for fully turbulent smooth-wall convection (solid line, Fig.~\ref{fig:NuRa}c) we see the scaling exponent is in the range $0.35\lesssim \Sca\lesssim 0.42$.
 The transitionally rough regime has a much larger exponent, with $0.45\lesssim\Sca\lesssim0.58$ for varying $k_s/L$ ratios and $Ra$. While the behavior of this exponent with Rayleigh number is dependent on the functional forms of $\Delta U_{fit}^+$ and $\Delta \Theta_{fit}^+$, the important aspect is that the use of these fits shows an exponent close to 0.55. As the flow enters the fully rough regime and $\Delta \Theta^+$ becomes constant ($Ra\approx 10^9$ for $k_s/L=1.14\times10^{-1}$, black dashed line), the scaling exponent reduces to approximately $0.42\lesssim\Sca\lesssim0.45$, consistent with Fig.~\ref{fig:NuRa}(a), although still larger than that of the smooth-wall flow. 
 Crucially, from  Eqs.~(\ref{eqn:UbFR_C}--\ref{eqn:TbFR}) we see that the rough-wall exponent must  depend on the equivalent sand-grain roughness,  $k_s/L$, as well as the temperature difference, $\Delta \Theta_{FR}^+$, making it distinct from the smooth-wall scaling exponent.

\section{Discussion and conclusions}

While we have used simple curve fits to obtain $\Sca\approx0.55$ in the transitionally rough regime, they show how large scaling exponents in the transitionally rough regime can be obtained \citep{xie17,zhu17b,Tummers19}.
Echoing Ref.~\cite{zhu17b}, these large exponents
 do not indicate that the asymptotic ultimate regime has been obtained, as both viscous and thermal diffusivity effects are still important in the transitionally rough regime (referred to as Regime I in \cite{zhu17b}).
It is only once the flow enters the fully rough regime (Regime II in \cite{zhu17b}), when the skin-friction coefficient is constant, that viscous actions become negligible
and the scaling exponent reduces in value.
Critically, however, thermal diffusivity effects will always remain important, as they do in the smooth wall.
From  Eq.~\ref{eqn:TbFR}, we see the fully rough centerline temperature and hence Nusselt number scales with the logarithm of $Re$, indicating that only for asymptotically large $Ra$ does the heat transfer of rough-wall flows approach
the asymptotic ultimate regime.

The origin for this major difference between momentum transfer and heat transfer in rough-wall shear flow 
lies in the pressure drag, which dominates the momentum transfer, but whose analog is absent for the heat
transfer \cite{Owen63,Cebeci84,mac19}. It is this absence which leads to an effective scaling $Nu\sim Ra^{0.42}$ in the fully rough ultimate regime, rather than the upper bound exponent 1/2.  Our $Nu$ vs $Ra$ scaling prediction also seems to be consistent with recent rough-plate RB experiments in the G{\"o}ttingen U-Boat facility, which in the ultimate regime yield an exponent of 0.43 for $Pr = 0.8$ and $Ra\approx10^{13}$ (E.~Bodenschatz, private communication). 


We finally note that due to the limitations in fabrication
 every surface is rough to some degree ($k_s/L>0$).
  For example, the G{\"o}ttingen U-Boat system with $L = 2.24$ m uses either aluminum (HPCF-I) or copper (HPCF-II) top and bottom plates with average roughness heights of $R_a\approx1.6$ $\mu$m and 0.2 $\mu$m, respectively \cite{Ahl09b}. 
  At what Rayleigh number will this roughness become visible in the $Nu(Ra)$ relation? Unfortunately, such an estimate is very difficult as it strongly depends on the prefactors and exact
  values of the scaling exponents. With the assumption of our above analysis,
   these surfaces could only be considered as fully smooth until 
 $Ra\approx1\times10^{15}$ and $9\times10^{16}$ for the aluminum and copper plates, respectively,
 before becoming transitionally rough with an enhanced scaling exponent.
These numbers should be taken with utmost care, as, as mentioned above,  these estimates strongly depend on the coefficients in the $Re-Ra$ relationship assumed in Section~\ref{sect:scaling}, with the present coefficients ($Re = 0.7Ra^{0.5}$) assuming fully turbulent boundary layers.
 If we use  $Re = 0.346Ra^{0.443}$ from Grossmann--Lohse theory \cite{gro11}, then the corresponding critical Rayleigh numbers are $4\times10^{17}$ and $6\times10^{19}$, respectively.
To obtain either set of values,
the roughness is assumed to be hydrodynamically and thermodynamically smooth ($\Delta U^+=\Delta\Theta^+=0$) until $k_s^+=4$,
and that the plate surface  equivalent sand-grain roughness is that of sinusoidal roughness, $k_s\approx 4.1k$,
where the sinusoidal semi-amplitude is related to the mean roughness height by $k=2.46R_a$ \cite{Chan15}.
While there are uncertainties in the exact $Ra$ values above, they provide some indication of the level of surface smoothness required in laboratory experiments to ensure the smooth-wall scaling exponent in the ultimate regime is observed.
Regardless, as shown in Figs.~\ref{fig:CfCh} and \ref{fig:NuRa}, where the different line colors correspond to varying $k_s/L$, for sufficiently large $Ra$ the surfaces will inevitably cease to be dynamically smooth.

\acknowledgements{The authors would like to gratefully acknowledge the financial support of the Australian Research Council through a Discovery Project (DP170102595).}

\bibliography{roughNC_bib,literatur}

\end{document}